\documentclass[11pt]{article}
\usepackage{epsfig}
\usepackage[usenames]{color}
\usepackage{graphicx}
\usepackage{amsmath}
\newcommand{\sbar}[1]{\ooalign{\hfil/\hfil\crcr$#1$}}

\def\la{\langle}\def\ra{\rangle}
\def\be{\begin{eqnarray}}\def\ee{\end{eqnarray}}
\def\lsim{\mathrel{\rlap{\lower3pt\hbox{\hskip1pt$\sim$}}
     \raise1pt\hbox{$<$}}} 
\def\gsim{\mathrel{\rlap{\lower3pt\hbox{\hskip1pt$\sim$}}
     \raise1pt\hbox{$>$}}} 
\def\le{ \begin{array}{ll}}\def\re{\end{array}}

\def\lear{ \left( \begin{array}{cc}}\def\rear{\end{array} \right)}
\def\tr{\textnormal{tr}}

\def\le{ \left( \begin{array}{cc}}\def\re{\end{array} \right)}
\def\tr{\textnormal{tr}}

\def\bi{\bibitem}
\def\bchi{\bar{\chi}}

\newcommand{\Slash}[1]{\ooalign{\hfil/\hfil\crcr$#1$}}
\renewcommand{\thefootnote}{\fnsymbol{footnote}}
\begin{document}
\hfill \vbox{\hbox{}}
\begin{center}{\Large\bf Interplay between $\omega$-Nucleon Interaction \\
and Nucleon Mass in Dense Baryonic Matter}\\[0.8cm]
{ Won-Gi Paeng\footnote{\sf e-mail: wgpaeng0@hanyang.ac.kr}
}\\
{\em Department of Physics, Hanyang University, Seoul 133-791, Korea}

{ Hyun Kyu Lee\footnote{\sf e-mail: hyunkyu@hanyang.ac.kr}
}\\
{\em Department of Physics, Hanyang University, Seoul 133-791, Korea}

{ Mannque Rho\footnote{\sf e-mail: mannque.rho@cea.fr }
}\\
{\em Institut de Physique Th\'eorique, CEA Saclay, 91191 Gif-sur-Yvette c\'edex, France \&
\\Department of Physics, Hanyang University, Seoul 133-791, Korea}

{ Chihiro Sasaki\footnote{\sf e-mail: sasaki@fias.uni-frankfurt.de}
}\\
{\em Frankfurt Institute for Advanced Studies,
D-60438 Frankfurt am Main, Germany}

\end{center}
\vspace{0.2cm}
\centerline{\today}
\vspace{0.2cm}
\begin{center}{\Large\bf Abstract}
\end{center}

 The dilaton-limit fixed point and the scaling properties of hadrons in the close vicinity of the fixed point in dense baryonic matter uncovered in hidden local symmetry implemented with spontaneously broken scale symmetry are shown to reveal a surprisingly intricate interplay, hitherto unsuspected, between the origin of the bulk of proton mass and the renormalization-group flow of the $\omega$-nuclear interactions. This rends a theoretical support to the previous (phenomenologically) observed correlation between the dropping nucleon mass and the behavior of the $\omega$-nuclear interactions in dense matter described in terms of half skyrmions that appear at a density denoted $n_{1/2}$ in skyrmion crystals. The role of the $\omega$-meson degree of freedom in the source for nucleon mass observed in this paper is highly reminiscent of its important role in the skyrmion description of nucleon mass in hidden local symmetric theory.  One of the most notable novel results found in this paper is that the nucleon mass in dense baryonic medium undergoes a drop roughly linear in density up to a density (denoted $\tilde{n}$) slightly above nuclear matter density ($n_0$) and then stays more or less constant up to the dilaton limit fixed point. The possibility that we entertain is that $\tilde{n}$ coincides with or at least close to $n_{1/2}$. We note that this feature can be economically captured by the parity-doublet model for nucleons with the chiral-invariant mass $m_0\sim (0.7-0.8) m_N$. It is found in one-loop renormalization-group analysis with the Lagrangian adopted that while the $\rho$-NN coupling ``runs" in density, the $\omega$-NN coupling does not scale: it will scale at two-loop or higher-loop order, but at a slower pace, so it is more appropriate to say it ``walks" rather than runs.  The former implies a drastic change in the nuclear tensor forces, affecting, among others, the nuclear symmetry energy and the latter generating the stiffness of the EoS at density higher than that of normal nuclear matter.


\vfill

\pagebreak
\setcounter{footnote}{0}
\renewcommand{\thefootnote}{\arabic{footnote}}
\section{Introduction and Conclusion}
If hidden local symmetry (HLS for short) is assumed to hold in the vicinity of chiral restoration in dense baryonic matter and if $U(2)$ symmetry is a good flavor symmetry for the vector mesons $\rho$ and $\omega$ in medium as it is in free space,  it follows from RG (renormalization-group) considerations that the $U(2)$ hidden gauge coupling constant $g$ and the ``effective" V (vector-meson)-nucleon coupling constant $g_{VNN}$  will scale in density as the quark condensate $\la\bar{q}q\ra$ scales, and consequently as $\la\bar{q}q\ra\rightarrow 0$ in the chiral limit, both the $\rho$-nucleon and $\omega$-nucleon couplings  as well as their masses will go to zero~\cite{HY:PR}. This implies two dramatic effects in nuclear processes even slightly above nuclear matter density: (1) The suppression of the $\rho$-nuclear coupling will remove the $\rho$ -exchange tensor force and hence strongly affect the nuclear symmetry energy~\cite{LPR}; (2) the suppression of the $\omega$-nucleon coupling will soften short-range repulsion in nuclear interactions and hence make the baryonic matter collapse at moderate density~\cite{sasaki}. The first effect, when treated appropriately,  turns out to be consistent with the EoS of compact-star matter that involves central densities of $\sim (5-6)n_0$ (where $n_0$ is the normal nuclear matter density), in fact playing a crucial role for explaining, within the HLS framework, the recently discovered 2-solar mass neutron star~\cite{dongetal}. On the other hand, the second effect, if unmodified, would be disastrous for the stability of baryonic matter in the density regime relevant to compact stars.

It is the purpose of this paper to suggest how to avoid the disaster due to the second (negative) effect without affecting the first ({\em positive}) effect. The key observation is that the $U(2)$ symmetry, seemingly good in the matter-free vacuum, must break down in medium, and hence the properties of the isovector and isoscalar vector mesons behave markedly differently as density is increased. The key element in this is the origin of the bulk of proton mass that appears to have no direct link to chiral symmetry, its RG flow and its unsuspected association with the property of the $\omega$-NN interaction in dense medium. Our conclusion is that while the $\rho$ and $\omega$ masses tend to zero (in the chiral limit), perhaps not in the same way, as density is increased, the effective $\omega$-nuclear coupling ``walks" in contrast to the effective $\rho$-nuclear coupling that ``runs." This feature was indicated in the phenomenology of compact-star matter studied in \cite{dongetal}, and we show in this paper how that feature can be understood in the framework of HLS, e.g., vector manifestation and dilaton-limit fixed point.

The basic assumption that we make is that local field theory can be applied to dense baryonic matter up to the density relevant to the EoS of compact stars. We will not, however, require that it be valid all the way to the chiral transition density denoted $n_c$.  In fact, we will not address what happens precisely at $n_c$ but consider {\em approaching} it from below. This means that we will not be able to properly account for the possibility of {\em explicit} quark degrees of freedom in discussing the EoS.

We consider HLS Lagrangian that contains as relevant degrees of freedom, the pions and the lowest-lying vector mesons, $\rho$ and $\omega$. It may very well be that for realistic treatment, as stressed recently~\cite{maetal}, the infinite tower of hidden local symmetric vector mesons as indicated in holographic QCD models~\cite{SS} need to be incorporated and the nucleons should be generated from such a generalized HLS Lagrangian. In this work, we will take the simplified Lagrangian in which the infinite-tower is integrated out leaving only the lowest vector mesons $V=(\rho,\omega)$ -- in addion to pions, add baryon fields coupled to the mesons \`a la HLS and implement a scalar dilaton field $\chi$ reflecting  spontaneously broken scale symmetry linked to the QCD trace anomaly\footnote{The role of scalar fields in effective Lagrangians is highly  problematic in general and it is not at all obvious how to do this also in our case. We will however be guided by phenomenology in low-energy nuclear physics, namely, the EFT Lagrangian be treated at mean field with the parameters of the Lagrangian ``sliding" with the density of the background in the spirit defined in \cite{BR91}. This will be the guiding principle in what follows.}. To be precise in definition, we shall call the dilaton-implemented HLS Lagrangian with baryon fields ``dBHLS" while ``$\Slash{d}$BHLS" will stand for dilatonless baryon HLS Lagrangian.  ``HLS" will stand for the generic notion of hidden local symmetry as well as for meson-only theory (without dilaton).

Our objective is to access nuclear matter as well as denser baryonic matter with a single Lagrangian, i.e., dBHLS. At zero density, that is, in the matter-free space, $\Slash{d}$BHLS is ``gauge equivalent" to baryon nonlinear sigma model and can be formulated, with the inclusion of  chiral loops, to give a consistent chiral perturbation theory with baryons and vector mesons in a way parallel to HLS~\cite{HY:PR}. For this, scalar excitations, in principle, can be generated from BHLS by loops.  The QCD scalar $\bar{q}q$ is high-lying and hence does not figure in nonlinear sigma model in hadron dynamics at low-energy.  Scalar glueball excitations will figure for the QCD trace anomaly, but they are also too massive. Thus the role of the dialton $\chi$ in the vacuum is unclear and remains an unsolved problem. At tree order, however, the $\Slash{d}$BHLS encodes the current algebra, and the dilaton $\chi$ could be suitably interpreted in dBHLS as the lowest scalar excitation in low-energy pion dynamics.

In going to nuclear matter density, one possible approach could be to  do in-medium chiral perturbation theory with $\Slash{d}$BHLS as one does with non-linear sigma model. However this requires high-order loop calculations, and this has not been done yet. What one can do instead is to do the mean field calculation with dBHLS as suggested in \cite{songetal,song}. The rationale there is that  doing the mean-field with a chiral Lagrangian of the dBHLS type near the nuclear saturation density is equivalent to doing Landau Fermi-liquid fixed point theory provided the parameters of the Lagrangian are suitably scaled~\cite{songetal,song,friman-rho}. What is required is that near the Fermi-liquid fixed point, the scalar $\chi$ should be  (predominantly) a chiral scalar with a mass around 600 MeV.

Now given the dBHLS Lagrangian with the parameters sliding with density, normalized at the nuclear saturation density $n_0$, the question then is how to go to higher density close to  the chiral transition point $n_c$? This question was raised and answered in \cite{sasaki}. It involves what is called ``dilaton limit fixed point" (DLFP for short) -- defined precisely later --  introduced by Beane and van Kolck~\cite{beane}. The basic idea is as follows: While the dilaton $\chi$ should be dominated by chiral singlet component at low density near $n_0$ in a complicated configuration consisting of multiquark and glueballs,  as one approaches chiral restoration at $n_c$, one should recover a Gell-Mann-L\'evy (GML)-type $SU(2)_L\times SU(2)_R$linear sigma model with  a scalar $\sigma$ making up the fourth component of the chiral four-vector ($\pi_1,\pi_2,\pi_3,\sigma$)~\cite{GML}.

For what follows, we need to generalize the GML sigma model to a parity-doubled model with nucleons in mirror assignment. In the original GML model (with nucleons in standard assignment), spontaneous breaking of chiral symmetry (in the chiral limit)  {\em entirely} generates both the mass of the scalar $\sigma$ and the mass of the nucleon. Thus in the limit that the symmetry is restored, both the $\sigma$ and the nucleon should become massless. We will argue that while the $\sigma$ mass could go to zero to join the triplet pions, the nucleon mass need not be zero at the symmetry restoration. This feature can be captured by introducing parity-doubling in the baryon configurations. We shall call the former ``standard" baryons and the latter ``parity-doubled" baryons. As will be elaborated in detail below, within the EFT framework, the nucleon mass cannot decrease much as density increases without getting into conflict with nature. To be specific, we write the nucleon mass parameter in the Lagrangian as
\be
m_N=m_0+ \Delta m (\kappa)\label{mass}
\ee
where $m_0$ is a chirally invariant mass term, a constant independent of the chiral condensate $\kappa\propto \la\bar{q}q\ra$, and $\Delta$ is dynamically generated mass that tends to zero as $\kappa\rightarrow 0$. The mass term $m_0$ can appear in chiral Lagrangian without upsetting chiral symmetry provided parity-doublets are introduced~\cite{detar}. A substantial $m_0$, indicated in nuclear phenomenology~\cite{dongetal}, will be the key issue in this paper.

The dilaton limit fixed point (DLFP) arrived at in \cite{sasaki,paeng} in the parity-doublet baryon model dBHLS can be summarized as follows. As shown in \cite{sasaki} -- and will be recalled in detail below, the idea is to make a field re-parametrization so that dBHLS Lagrangian with chiral symmetry in nonlinear realization can be {\em linearized} in the limit $\kappa\rightarrow 0$.  If one assumes that $U(2)$ symmetry holds for $\rho$ and $\omega$, one arrives at the DLFP of the form
\be
(g_V-1,g_A-1, {\Delta m}\equiv m_N-m_0)\rightarrow (0,0,0) \ \ as \ \ \kappa\rightarrow 0\label{DLFP}
\ee
where $V=(\rho,\omega)$, $g_A$ is the axial vector coupling constant and $g_V$ is the ``induced" $V$-nucleon coupling defined by the effective vector-meson-nucleon coupling
\be
g_{VNN}=g(g_V-1)\label{gvnn}
\ee
where $g$ is the $U(2)$ hidden gauge coupling constant. As $\kappa\rightarrow 0$, we expect that, even for $g\neq 0$,
\be
g_{\rho NN}=g_{\omega NN}\rightarrow 0.\label{dlfpu2}
\ee
The ``vector manifestation" (VM) of HLS (and $\Slash{d}$BHLS) corresponds to $g\rightarrow 0$. We can see from (\ref{gvnn}) and (\ref{dlfpu2}) that the ${\rho NN}$ coupling $g_{\rho NN}$ can go to zero before the VM/HLS fixed point (identified with chiral restoration) is reached.

Several important consequences follow from the property (\ref{dlfpu2}).

If dBHLS were applied in the mean-field approximation to  baryonic matter much denser than nuclear matter, then one would expect that the nuclear symmetry energy that figures importantly in the EoS for compact-star matter, proportional to $g_{\rho NN}^2$, decreases and vanishes at the DLFP. One should however note that there is no solid argument why the mean-field approximation should hold at density much higher than the saturation density, i.e., Fermi-liquid fixed point, where mean-field approximation is valid. In fact, $g_{\rho NN}\rightarrow 0$ would imply that the $\rho$ contribution to the nuclear tensor forces get suppressed, leaving the pion tensor force more effective. This feature turns out to lead to the stiffening -- instead of the softening -- of the symmetry energy, contrary to the mean-field expectation, as has been observed in effective field theory approach to nuclear dynamics. Indeed what was found in \cite{dongetal} provides a rather strong support for the prediction (\ref{dlfpu2}) for the $\rho$ meson.  This suggests that the mean-field approximation breaks down at a density above nuclear matter. These matters are further discussed in Sec.\ref{SE}. { As mentioned therein, this implies that associated with the skyrmion-half-skyrmion topological transition that plays an important role in the calculation of EoS in \cite{dongetal}, changeover takes place from a Fermi liquid structure to a non-Fermi liquid structure, resembling certain quantum critical phenomena in condensed matter.}

The situation with the $\omega$-NN coupling is quite different. As noted in \cite{sasaki}, (\ref{dlfpu2}) would imply the suppression of the principal mechanism in EFT involving $\omega$ exchanges for the short distance repulsion in nuclear interactions indispensable at high density. Since one cannot pinpoint the density at which the DLFP sets in, one cannot say that the predicted suppression of the $\omega$-exchange repulsion is inconsistent with nature. However it is generally believed in nuclear community that any significant reduction of the $\omega$-NN coupling would make the EoS of nuclear matter too soft, causing difficulty in getting correct saturation. In \cite{dongetal}, it was observed that unless the nucleon mass is suitably reduced, thereby increasing repulsion, it would be difficult to reduce the $\omega$-NN coupling. This suggests a close correlation between the behavior of the nucleon mass and the $\omega$-NN coupling in dense medium. This issue will be further elaborated on in Sec.~\ref{repulsion}. One could investigate the  interplay between the two phenomenologically in the EFT formalism employed in \cite{dongetal}. In this paper, we will present a theoretical reasoning as to how the interplay can take place.

Taking the hint that the $U(2)$ symmetry can be substantially broken in medium, what we propose is to depart from hidden local symmetry in $[U(2)_L\times U(2)_R]_{\rm global}\times [U(2)_V]_{\rm local}$ and consider $ [SU(2)_L\times SU(2)_R]_{\rm global}\times [SU(2)_V\times U(1)_V]_{\rm local}$. Denoting gauge couplings by $g_{\rho}$ and $g_{\omega}$ and induced couplings by $g_{V\rho}$ and $g_{V\omega}$, an analysis parallel to the $[U(2)]_{local}$ case~\cite{sasaki,paeng} reveals that as $\kappa\rightarrow 0$,  one approaches the DLFP
\be
(g_{V\rho}-1,g_A-g_{V\rho}, {\Delta m})\rightarrow (0,0,0),
\ee
which is what was found in \cite{sasaki,paeng}. But there is a major difference for the $g_{V\omega}-1$ coupling: There is no constraint for that coupling in going toward the DLFP. Furthermore as will be shown below,  one-loop RG analysis shows that $g_{V\omega}$ does not ``run" in contrast to $g_{V\rho}$ which drops rapidly to zero toward the DLFP. This leads to our conclusion that as density increases above $n_0$, the $U(2)$ symmetry must be broken down significantly, and the $g_{\omega NN}$ coupling {\em does not drop as fast as   $g_{\rho NN}$ does}.

In short, the RG analysis with $\Slash{d}$BHLS and the mean field treatment with dBHLS, both given in this paper, the large $N_c$ analysis of dense skyrmion matter described in \cite{LR} and the phenomenological study of the EoS for compact-star matter of \cite{dongetal} all converge to the conclusion that the dropping of the in-medium nucleon mass stops at $\tilde{n}\approx n_{1/2}\sim 2n_0$ with the mass staying constant up to near chiral restoration, suggesting an $m_0\sim (0.7-0.8) m_N$ in the parity-doublet model for baryons. This property is found to have an intimate connection with the role that the $\omega$-meson degree of freedom plays in the structure of nucleon and nuclear matter.

In what follows, we provide details to what are given in sketch above.

\section{Hidden Local Symmetric Parity-Doublet Model}

In this section,  we give a precise definition of the model we will study. We will first write down and discuss the hidden local symmetric parity doublet model without dilaton field, i.e., $\Slash{d}$BHLS~\cite{paeng}.  We will be focusing on the case where the chiral invariant mass $m_0$ in (\ref{mass}) is non-vanishing, in fact, substantially big, but we will also discuss the case for $m_0=0$, i.e.,  ``standard" scenario.

We will limit our considerations to two flavors ($N_f =2$).
Motivated by the finding in \cite{dongetal}, we relax the $U(2)_V$ symmetry
for $\rho$ and $\omega$ mesons. We assume that the  symmetry of the Lagangian involved is  $G_{\rm{global}} \times H_{\rm{local}}$, where $G_{\rm global}=[SU(2)_L \times SU(2)_R]_{\rm global}$ is the global chiral symmetry and $H_{\rm local}=[SU(2)_V \times U(1)_V]_{\rm local}$ is the hidden local symmetry. We take $\rho$ as the gauge bosons of
$[SU(2)_V]_{\rm local}$ and $\omega$ as of $[U(1)_V]_{\rm local}$.
The basic quantities are the HLS gauge bosons, $V_\rho^\mu$ and $V_\omega^\mu$,
\begin{eqnarray}
V_\rho^\mu = g_\rho \vec{\rho}^{\,\mu} \cdot \frac{\vec{\tau}}{2}\,, \\
V_\omega^\mu = g_\omega \frac{\omega^\mu}{2}
\end{eqnarray}
where $g_\rho$ and $g_\omega$ are the HLS gauge couplings that will be taken unequal for the local $SU(2)\times U(1)$ symmetry concerned. The vectors transform
\begin{eqnarray}
V_\rho^\mu \to h\,V_\rho^\mu\, h^\dagger - i\partial^\mu h\cdot h^\dagger\,, \\
V_\omega^\mu \to u\,V_\omega^\mu\, u^\dagger - i\partial^\mu u\cdot u^\dagger\,,
\end{eqnarray}
with $h=h(\pi(x), g_L, g_R) \in \left[ SU(2)_V\right]_{\rm local}$
and $u=u(\pi(x), g_L, g_R) \in \left[ U(1)_V\right]_{\rm local}$
in terms of $g_{L,R} \in \left[ SU(2)_{L,R}\right]_{\rm global}$
and  the two matrix valued variables
$\xi_L$ and  $\xi_R$, combined in a $2 \times 2$ special-unitary matrix
representing the pion field
\be
U = \xi_L^\dagger \xi_R=e^{2i\pi/F_\pi}
\ee
transforming
\begin{equation}
U \to g_L U g_R^\dagger\,.
\end{equation}
The variables $\xi$'s transform as
\begin{equation}
\xi_{L,R} \to u h\,\xi_{L,R}\,g_{L,R}^\dagger.
\end{equation}
They may be parameterized as
\begin{eqnarray}
\xi_{L,R}= e^{\frac{i}{2}\sigma_\omega/{F_{\sigma\omega}}} e^{i\sigma_\rho/{F_{\sigma\rho}}}e^{\mp i\pi/{F_\pi}}.
\end{eqnarray}
Here $\pi = \pi^a\frac{\tau_a}{2}$ denote the pseudoscalar Nambu-Goldstone (NG)
bosons associated with the spontaneous breaking of $G_{\rm{global}}$ chiral symmetry, and $\sigma_\omega$ and $\sigma_\rho (= \sigma_\rho^a \frac{\tau_a}{2})$ are the NG bosons associated with the spontaneous symmetry breaking of $U(1)_{\rm local}$ and $SU(2)_{\rm{local}}$ respectively. The $\sigma$s are absorbed into the HLS gauge bosons through the Higgs mechanism, giving rise to their HLS boson masses. $F_\pi$,  $F_{\sigma\omega}$ and $F_{\sigma\rho}$ are decay constants of the associated particles.

To construct hidden local symmetric Lagrangian, it is convenient to introduce the Maurer-Cartan 1-forms
\begin{eqnarray}
\hat{\alpha}_{\perp }^{\mu}
&=& \frac{1}{2i}\left[ D^\mu\xi_R \cdot \xi_R^{\dagger}
{}- D^\mu\xi_L \cdot \xi_L^{\dagger} \right]\,,
\nonumber\\
\hat{\alpha}_{\parallel}^{\mu}
&=& \frac{1}{2i}\left[ D^\mu\xi_R \cdot \xi_R^{\dagger}
{}+ D^\mu\xi_L \cdot \xi_L^{\dagger} \right]\,,
\end{eqnarray}
with the covariant derivatives of $\xi_{L,R}$
\begin{eqnarray}
&&
D^\mu \xi_L
 = \partial^\mu\xi_L - iV_\rho^\mu\xi_L - iV_\omega^\mu\xi_L\,,
\\
&&
D^\mu \xi_R
 = \partial^\mu\xi_R - iV_\rho^\mu\xi_R - iV_\omega^\mu\xi_R\,.
\end{eqnarray}
transforming homogeneously,
\begin{equation}
\hat{\alpha}_{\perp,\parallel}^\mu
\to u\, h\, \hat{\alpha}_{\perp,\parallel}^\mu\, h^\dagger\, u^\dagger \,,
\end{equation}
and $\hat{\alpha}_\perp \in {\mathcal G}-{\mathcal H}$ and
$\hat{\alpha}_\parallel \in {\mathcal H}$.

With the above definitions, we can immediately write down the Lagrangian for the mesonic sector.  To the leading order in derivative expansion (i.e., to ${\cal O}(p^2)$), it is
\begin{eqnarray}
{\mathcal L}_M
& = & F_\pi^2\mbox{tr}\left[ \hat{\alpha}_{\perp\mu}
  \hat{\alpha}_{\perp}^{\mu} \right]
{}+ F_{\sigma\rho}^2\mbox{tr}\left[ \hat{\alpha}_{\parallel\mu}
  \hat{\alpha}_{\parallel}^{\mu} \right]
{}+ F_{0}^2\mbox{tr}\left[ \hat{\alpha}_{\parallel\mu} \right]
  \mbox{tr}\left[ \hat{\alpha}_{\parallel}^{\mu} \right] \nonumber \\
&& {}- \frac{1}{2g_\rho^2}\mbox{tr}\left[ V_{\mu\nu}V^{\mu\nu} \right]
{}- \frac{1}{2g_0^2}\mbox{tr}\left[ V_{\mu\nu}\right]\mbox{tr}\left[V^{\mu\nu} \right]\,,
\label{lagmeson}
\end{eqnarray}
with
\be
 F_{0}^2 &=& \frac{F_{\sigma\omega}^2 - F_{\sigma\rho}^2}{2},\nonumber\\
   \frac{1}{g_0^2} &=& \frac{1}{2}\left(\frac{1}{g_\omega^2} - \frac{1}{g_\rho^2} \right)
   \ee
and the field strength is given by
\begin{eqnarray}
& V^{\mu\nu} = \partial^\mu V^\nu - \partial^\nu V^\mu
{}- i\left[ V^\mu, V^\nu \right]\,, \\
& V^{\mu} = V_\rho^\mu + V_\omega^\mu \,.
\end{eqnarray}
We recover local $U(2)$ symmetry if we set $F_{\sigma\omega}^2 = F_{\sigma\rho}^2$ and $g_\omega=g_\rho=g$ in (\ref{lagmeson}).

{Now, we construct the Lagrangian with nucleons in a hidden local symmetric model.

When the nucleon transforms as
\begin{eqnarray}
\psi_{1,\,L} &\equiv& P_L \psi_1 \rightarrow g_L P_L \psi_1\,,\label{ln1lm}\\
\psi_{1,\,R} &\equiv& P_R \psi_1 \rightarrow g_R P_R \psi_1\,,\label{ln1rm}
\end{eqnarray} under chiral transformation in the Gell-Mann-L\'evy-type linear sigma model,
we assign the following transformation,
\begin{eqnarray}
\psi_{2,\,L} &\equiv& P_L \psi_2 \rightarrow g_R P_L \psi_2\,,\label{ln2lm}\\
\psi_{2,\,R} &\equiv& P_R \psi_2 \rightarrow g_L P_R \psi_2\,,\label{ln2rm}
\end{eqnarray} to the nucleon's chiral partner in the mirror assignment\cite{detar},
where $P_{L,\,R} = \frac{1\mp \gamma_5}{2}$.
In hidden local symmetry model in the mirror assignment, the nucleon and its chiral partner
are represented in the non-linearized form, $Q$, under chiral transformation, which is given as a function of $\xi_{L,R}$ and $\psi_{1,2}$,
\begin{equation}
Q \equiv Q\left( \xi_{L,R},\ \psi_{1,2}\right)\,,
\end{equation} with the nucleon doublet $Q = \left(\begin{array}{cc} Q_{1} \\ Q_{2} \end{array}\right)$
transforming as
\begin{equation}
Q \to u\,h\,Q\,
\end{equation} under hidden local transformation.
}

One can readily write down, following \cite{sasaki,paeng}, the  Lagrangian for parity-doublet nucleons coupled to HLS vectors
\begin{eqnarray}
\mathcal{L}_{N}
& = & \bar{Q}i\gamma^{\mu}D_{\mu}Q - g_{1}F_{\pi}\bar{Q}Q
{}+ g_{2}F_{\pi}\bar{Q}\rho_{3}Q
{}- im_{0}\bar{Q}\rho_{2}\gamma_{5}Q
\nonumber\\
&&+ g_{V\rho} \bar{Q}\gamma^{\mu}\hat{\alpha}_{\parallel \mu}Q
{}+ g_{V0} \bar{Q}\gamma^{\mu}\mbox{tr}\left[\hat{\alpha}_{\parallel \mu} \right]Q
{}+ g_{A}\bar{Q} \rho_{3} \gamma^{\mu}\hat{\alpha}_{\perp \mu}
\gamma_{5} Q\,,
\label{lagnucleon}
\end{eqnarray}
where the covariant derivative of Q is
\be
D_\mu Q = \left( \partial_\mu - iV_{\rho\mu} - iV_{\omega\mu} \right) Q,
\ee
and the $\rho_{i}$ are the Pauli matrices acting on the parity-doublet.
$g_A$, $g_{V\rho}$ and $g_{V0}\equiv \frac 12 (g_{V\omega}-g_{V\rho})$ are dimensionless parameters.
To diagonalize the mass term in Eq.~(\ref{lagnucleon}), we transform $Q$ into
a new field $N$:
\begin{eqnarray}
\left( \begin{array}{cc} N_{+} \\ N_{-} \end{array} \right)
= \frac{1}{\sqrt{2 \cosh \delta}} \left( \begin{array}{cc} e^{\delta/2} & \gamma_{5}e^{-\delta/2}
\\ \gamma_{5}e^{-\delta/2} & -e^{\delta/2} \end{array} \right)
\left( \begin{array}{cc} Q_{1} \\ Q_{2} \end{array} \right)\,,
\end{eqnarray}
where {$\sinh \delta =  \frac{g_{1}F_{\pi}}{m_{0}}$}.
We identify $N_{\pm}$ as parity-even and parity-odd states respectively.
The nucleon masses are found to be
\begin{eqnarray}
&& m_{N_{\pm}}
= \mp  g_{2} F_{\pi} + \sqrt{\left( g_{1}F_{\pi}\right)^{2} + m_{0}^{2}}\,,
\label{mass}
\\
&& \cosh \delta = \frac{m_{N_+} + m_{N_-} }{2m_{0}}\,.
\end{eqnarray}
Finally, we arrive at the Lagrangian in parity eigenstate as
\begin{eqnarray}
\mathcal{L}_{N}
&=& \bar{N} i \sbar{D} N - \bar{N}  \hat{\mathcal{M}} N
{}+ g_A\bar{N}\gamma^\mu\hat{G}\hat{\alpha}_{\perp\mu}\gamma_5 N
\nonumber \\
&&
{}+ g_{V\rho}\bar{N}\gamma^{\mu} \hat{\alpha}_{\parallel\mu} N
{}+ g_{V0} \bar{N}\gamma^{\mu}\mbox{tr}\left[\hat{\alpha}_{\parallel \mu} \right]N\,,
\label{Nlagrangian}
\\
\hat{\mathcal{M}}
&=& \left( \begin{array}{cc} m_{N_+} & 0 \\
0 & m_{N_-} \end{array} \right)\,,
\quad
\hat{G} = \left( \begin{array}{cc}
\tanh\delta & \gamma_5/\cosh\delta \\
\gamma_5/\cosh\delta & -\tanh\delta
\end{array}\right)\,.
\end{eqnarray}

It is convenient for the analysis of $\omega$-nucleon coupling to change slightly the Lagrangian (\ref{Nlagrangian}). We define quantities belonging to
the algebra of {$G_{\rm global} \times [SU(2)_V]_{\rm local}$} as
\begin{eqnarray}
& \tilde{ \alpha}^\mu_\perp = \frac{1}{2i} \left[ {\cal D}^\mu \tilde{ \xi}_R \cdot \tilde{ \xi}_R^{\dagger} - {\cal D}^\mu \tilde{ \xi}_L \cdot \tilde{ \xi}_L^{\dagger}  \right], \\
& \tilde{ \alpha}^\mu_\parallel = \frac{1}{2i} \left[ {\cal D}^\mu \tilde{ \xi}_R \cdot \tilde{ \xi}_R^{\dagger} + {\cal D}^\mu \tilde{ \xi}_L \cdot \tilde{ \xi}_L^{\dagger}  \right],
\end{eqnarray} where
\begin{eqnarray}
\tilde{\xi}_{L,R}= e^{i\sigma_\rho/{F_{\sigma\rho}}}e^{\mp i\pi/{F_\pi}}\,,
\end{eqnarray} and
\begin{eqnarray}
&&
{\cal D}^\mu \tilde{ \xi}_L
 = \partial^\mu\tilde{ \xi}_L - iV_\rho^\mu\tilde{ \xi}_L\,,
\nonumber\\
&&
{\cal D}^\mu \tilde{ \xi}_R
 = \partial^\mu\tilde{ \xi}_R - iV_\rho^\mu\tilde{ \xi}_R\,.
\end{eqnarray}
{$\hat{\alpha}$ and $\tilde{\alpha}$ are related to each other via
\begin{eqnarray}
\hat{\alpha}^\mu_\perp &=& \tilde{\alpha}^\mu_\perp\,, \\
\hat{\alpha}^\mu_\parallel &=& \tilde{\alpha}^\mu_\parallel +\frac{\partial^\mu \sigma_\omega}{2 F_{\sigma \omega}} - g_{\omega} \frac{\omega^\mu}{2}\,.
\end{eqnarray}}
Then, the Lagrangians (\ref{lagmeson}) and (\ref{Nlagrangian}) take the form
\begin{eqnarray}
{\mathcal L}_M
& = & F_\pi^2\mbox{tr}\left[ \tilde{\alpha}_{\perp\mu}
  \tilde{\alpha}_{\perp}^{\mu} \right]
{}+ F_{\sigma\rho}^2\mbox{tr}\left[ \tilde{\alpha}_{\parallel\mu}
  \tilde{\alpha}_{\parallel}^{\mu} \right]
{}+ \frac{F_{\sigma\omega}^2}{2} \left( \frac{\partial_\mu\sigma_\omega}{F_{\sigma\omega}} - g_\omega \omega_\mu  \right)
   \left( \frac{\partial^\mu\sigma_\omega}{F_{\sigma\omega}} - g_\omega \omega^\mu  \right) \nonumber \\
&& {}- \frac{1}{2}\mbox{tr}\left[ \rho_{\mu\nu}\rho^{\mu\nu} \right]
{}- \frac{1}{2}\mbox{tr}\left[ \omega_{\mu\nu}\omega^{\mu\nu} \right]\,,
\label{lagmeson2}
\end{eqnarray}
\begin{eqnarray}
\mathcal{L}_{N}
&=& \bar{N} i \sbar{D} N - \bar{N}  \hat{\mathcal{M}} N
{}+ g_A\bar{N}\gamma^\mu\hat{G}\tilde{\alpha}_{\perp\mu}\gamma_5 N
\nonumber \\
&&
{}+ g_{V\rho}\bar{N}\gamma^{\mu} \tilde{\alpha}_{\parallel\mu} N
{}+ g_{V\omega} \bar{N}\gamma^{\mu}\left( \frac{ \partial_\mu \sigma_\omega}{2F_{\sigma\omega}} - g_\omega \frac{\omega_\mu}{2}  \right)N\,,
\label{Nlagrangian2}
\end{eqnarray}
where $g_{V0}$ is replaced by $\frac 12 (g_{V\omega}-g_{V\rho})$
{and
\begin{eqnarray}
& \rho^{\mu\nu} = \partial^\mu \vec{\rho}^{\,\nu} \cdot \frac{\vec{\tau}}{2} - \partial^\nu \vec{\rho}^{\,\mu} \cdot \frac{\vec{\tau}}{2}
{}- ig_\rho \left[ \vec{\rho}^{\,\mu} \cdot \frac{\vec{\tau}}{2}, \vec{\rho}^{\,\nu} \cdot \frac{\vec{\tau}}{2} \right]\,, \\
& \omega^{\mu\nu} = \partial^\mu \frac{\omega^\nu}{2} - \partial^\nu \frac{\omega^\mu}{2}\,.
\end{eqnarray}}
Note that the $\omega$ meson couples to nucleon and NG $\sigma_\omega$, but there is no $\omega$ coupling to other mesons, i.e.,  $\pi$ and $\rho$, at tree order. There can be tree-order $\omega\pi^3$ and $\omega$-$\pi$-$\rho$ couplings in the homogeneous Wess-Zumino term in the anomalous part of the HLS Lagrangian that could give rise to a one-loop correction to $\omega$-nucleon coupling, but does not contribute at the order we are working with. We willl see in Sec.~\ref{RG} that at the one-loop order, the coupling $g_{V\omega}$ does not scale.

One can read off the vector meson mass and the $\rho\pi\pi$ coupling constant at tree level as
\begin{eqnarray}
&&
m_\rho^2 = g_\rho^2 F_{\sigma \rho}^2\,,\label{rhomass}
\quad
m_\omega^2 = g_\omega^2 F_{\sigma \omega}^2\,,\label{omegamass}
\\
&&
g_{\rho\pi\pi} = \frac{1}{2}a_\rho g_\rho\,,
\\
&&
a_\rho = \frac{F_{\sigma\rho}^2}{F_\pi^2}\,.
\end{eqnarray}
The vector mesons couple to nucleons as
\begin{eqnarray}
g_{\rho N_+ N_+} & = & g_{\rho N_- N_-} = \left( g_{V \rho } - 1  \right) g_\rho\,,
\\
g_{\omega N_+ N_+} & = & g_{\omega N_- N_-} = \left( g_{V\omega} - 1 \right) g_\omega\,
\end{eqnarray}
and the axial vectors coupling as
\begin{eqnarray}
g_{AN_{+}N_{+}} = - g_{AN_{-}N_{-}} = g_{A} \tanh \delta\,
\end{eqnarray}
where the subscripts stand for the parity of the nucleon doublet. When $U(2)$ symmetry is restored, we will have $g_{\omega NN} = g_{\rho NN}$ and the $\omega$-nuclear coupling will vanish as in \cite{sasaki} in approaching the DLFP. In what follows, we will not assume $U(2)$ symmetry.

\section{Going Towards the Dilaton-Limit Fixed Point \label{dBHLS}}

In order to study what happens to the baryonic matter as density increases, we need to incorporate the dilaton field $\chi$ that represents spontaneously broken scale symmetry of QCD. The explicit scale symmetry breaking associated with the trace anomaly that is also  presumably responsible for the spontaneous breaking~\cite{LR} will not figure directly in our consideration.

We follow the standard trick of inserting the ``conformal compensator" field $\chi$ into the Lagrangian (\ref{lagmeson}) and (\ref{lagnucleon}) to  obtain scale symmetric Lagrangian, with the scale invariance broken spontaneously. It is given by
\begin{eqnarray}
{\cal L} &=& {\cal L}_N + {\cal L}_M + {\cal L}_\chi\,,
\label{dlfplag} \\
\mathcal{L}_{N}
&=& \bar{Q}i\gamma^{\mu}D_{\mu}Q - g_{1}F_{\pi}\frac{\chi}{F_{\chi}}\bar{Q}Q
{}+ g_{2}F_{\pi}\frac{\chi}{F_{\chi}}\bar{Q}\rho_{3}Q
{}- im_{0}\bar{Q}\rho_{2}\gamma_{5}Q
\nonumber\\
&&+ g_{V\rho} \bar{Q}\gamma^{\mu}\hat{\alpha}_{\parallel \mu}Q
{}+ g_{V0} \bar{Q}\gamma^{\mu}\mbox{tr}\left[\hat{\alpha}_{\parallel \mu} \right]Q
{}+ g_{A}\bar{Q} \rho_{3} \gamma^{\mu}\hat{\alpha}_{\perp \mu}
\gamma_{5} Q\,,
\label{NchiLargrangian} \\
{\mathcal L}_M
& = & \frac{F_{\pi}^2}{F_{\chi}^2} \chi^2\mbox{tr}\left[ \hat{\alpha}_{\perp\mu}
  \hat{\alpha}_{\perp}^{\mu} \right]
{}+ \frac{F_{\sigma\rho}^2}{F_{\chi}^2} \chi^2\mbox{tr}\left[ \hat{\alpha}_{\parallel\mu}
  \hat{\alpha}_{\parallel}^{\mu} \right]
{}+ \frac{F_{\sigma\omega}^2 - F_{\sigma\rho}^2}{2F_{\chi}^2} \chi^2\mbox{tr}\left[ \hat{\alpha}_{\parallel\mu} \right]
  \mbox{tr}\left[ \hat{\alpha}_{\parallel}^{\mu} \right] \nonumber \\
&& {}- \frac{1}{2}\mbox{tr}\left[ \rho_{\mu\nu}\rho^{\mu\nu} \right]
{}- \frac{1}{2}\mbox{tr}\left[ \omega_{\mu\nu}\omega^{\mu\nu} \right]\,,
 \\
{\mathcal L}_\chi
&=& \frac{1}{2}\partial_\mu\chi\cdot\partial^\mu\chi {}-V(\chi)
\end{eqnarray} where $V(\chi)$ is the Coleman-Weinberg-type dilaton potential that breaks scale symmetry spontaneously.
We do not write down its explicit form since it is not needed for our purpose.
Here, $F_{\chi}$ is the vacuum expectation value of $\chi$ at zero temperature and density.

To move towards a chiral symmetric GML-type linear sigma model, we do the field re-parametrizations $\Sigma=U\chi\frac{F_\pi}{F_\chi}=s+i\vec{\tau}\cdot \vec{\pi}$ -- that also defines scalar $s$ -- and
\begin{equation}
\psi_{1,2}
= \frac{1}{2}\left[ \left( \xi_R^{\dagger}+ \xi_L^{\dagger} \right)
\pm \gamma_5\left( \xi_R^{\dagger} - \xi_L^{\dagger}  \right) \right] Q_{1,2}\, ,\label{e6}
\end{equation} or equivalently
\begin{eqnarray}
\psi & = &  \frac{1}{2}\left[ \left( \xi_R^{\dagger}+ \xi_L^{\dagger} \right)
+ \rho_{3} \gamma_5 \left( \xi_R^{\dagger} - \xi_L^{\dagger}  \right) \right] Q, \\
Q & = &  \frac{1}{2} \left[ \left( \xi_R + \xi_L \right)
+ \rho_{3} \gamma_5 \left( \xi_R - \xi_L  \right) \right] \psi\,.\label{e2}
\end{eqnarray}
With these reparametrized fields and going to parity eigenstates, one finds a complicated expression for (\ref{dlfplag}) composed of a part that is regular, ${\cal L}_{\rm reg}$, and a part that is singular, ${\cal L}_{\rm sing}$, as $\mbox{tr}(\Sigma\Sigma^\dagger)\equiv\kappa^2 = 2\left( s^2 + \pi^{a\,2}\right)
\rightarrow 0$, where $a$ is iso-spin index.
The singular part that arises solely from the scale invariant part of the
original Lagrangian (\ref{dlfplag}) has the form
\begin{eqnarray}
\mathcal{L}_{\rm sing} =
\left( g_{V\rho} -g_A \right) {\cal A} \left( 1/\tr \left[ \Sigma \Sigma^{\dagger} \right]\right) + \left( \alpha -1\right) {\cal B} \left( 1/\tr \left[ \Sigma \Sigma^{\dagger} \right]\right)\,,
\label{sing}
\end{eqnarray}
where $\alpha \equiv \frac{F_\pi^2}{F_\chi^2}$ and
\be
{\mathcal A}
&=&
\frac{ -i }{4} \tr \left( \Sigma \Sigma^{\dagger} \right)^{-2} \bar{\psi} \left[  \tr\left( \sbar{\partial} \left(\Sigma \Sigma^{\dagger} \right)  \right)\left\{ \Sigma, \Sigma^{\dagger} \right\} - 2 \tr\left( \Sigma \Sigma^{\dagger} \right) \left( \Sigma \sbar{\partial} \Sigma^{\dagger} + \Sigma^{\dagger} \sbar{\partial} \Sigma \right) \right] \psi \nonumber \\
&& \frac{ -i }{2} \tr\left( \Sigma \Sigma^{\dagger} \right)^{-1} \bar{ \psi } \rho_{3} \gamma_{5} \left( \Sigma \sbar{\partial} \Sigma^{\dagger} - \Sigma^{\dagger} \sbar{\partial} \Sigma \right) \psi \\
{\mathcal B} &=& \frac{-1}{16 \alpha } \mbox{tr} \left( \Sigma \Sigma^\dagger \right)^{-1} \mbox{tr} \left[ \partial_\mu \left( \Sigma \Sigma^\dagger \right)\right] \mbox{tr} \left[ \partial^\mu \left( \Sigma \Sigma^\dagger \right)\right] \,.
\ee
That ${\mathcal L}_{\rm sing}$ be absent leads to the conditions that

\be
g_{V\rho}-g_A\rightarrow 0\,,
\quad
\alpha -1 \to 0\,.
\ee
Using large $N_c$ sum-rule arguments~\cite{beane}
and the RGE given in the next section
, we infer\footnote{
It is most plausible that $g_A - 1 = 0$ is reached only after $g_A - g_{V\rho} = 0$ is reached. This is because $\mbox{tr}(\Sigma\Sigma^\dagger) \rightarrow 0$ gives the constraint $g_A - g_{V\rho}=0$, but not $g_A - 1 = 0$.  However the RGE given in the next section has the infrared fixed point $g_{V\rho}-1=0$, so the point at which $g_A - g_{V\rho} = 0$ could be very near the chiral restoration point at which $g_A$ is close to 1. In fact it is observed phenomenologically in Gamow-Teller transitions in heavy nuclei that $g^*_A\approx 1$.}
\be
g_A-1\rightarrow 0\,.
\ee
In the density regime where GML-type linear sigma model is valid, the nucleon mass can be given as
\begin{equation}
m_{N_\pm} = \mp g_2 \langle s \rangle + \sqrt{\left( g_1 \langle s \rangle \right)^2 +  m_0^2}\,,\label{nmass}
\end{equation} where $\langle s \rangle$ is the vacuum expectation value of $s$.
As the chiral symmetry restoration point is approached, $\langle s \rangle\rightarrow 0$, so in the limit $\mbox{tr}(\Sigma\Sigma^\dagger) \rightarrow 0$,
we expect
\begin{equation}
m_{N_\pm} \rightarrow m_0\,.
\end{equation}
These are the constraints that lead to the dilaton limit as in \cite{sasaki} and announced above. It follows then that
\be
g_{\rho NN}=g_\rho(g_{V\rho}-1)\rightarrow 0.
\ee
We thus find that in the dilaton limit, the $\rho$ meson decouples from the nucleon. In contrast, the limiting $\mbox{tr}(\Sigma\Sigma^\dagger)\rightarrow 0$ {\em does not} give any constraint on $(g_{V\omega}-1)$. The $\omega$-nucleon coupling remains non-vanishing in the Lagrangian which in  unitary gauge (with $\sigma_{\omega}=\sigma_\rho=0$) and in terms of fluctuations $\tilde{s}$ and $\tilde{\pi}$
around their expectation values, takes the form
\begin{eqnarray}
{\mathcal L}_N
=
&\bar{\cal N}i\sbar{\partial}{\cal N} - \bar{\cal N}\hat{M}{\cal N}
{}- g_1\bar{\mathcal N}\left(
\hat{G}\tilde{s} + \rho_3\gamma_5 i\vec{\tau}\cdot\vec{\tilde{\pi}}
\right) {\mathcal N}
\nonumber\\
&
{}+ g_2\bar{\mathcal N}\left(
\rho_3 \tilde{s} + \hat{G}\gamma_5 i\vec{\tau}\cdot\vec{\tilde{\pi}}
\right) {\mathcal N}
+ \left(1-g_{V\omega} \right) g_\omega{\cal N}  \frac{\sbar{\omega}}{2}  \mathcal{N}\,,
\end{eqnarray}
which is the same as the Lagrangian given in \cite{detar} except for the $\omega$-nucleon interaction. This is just the nucleon part of the linear sigma model in which the $\omega$ is minimally coupled to the nucleon, applicable infinitesimally below the critical density $n_c$ with the mass $\hat{M}$ replacing $m_0$. What is significant with this result is that it shows that the suppression of the $\omega$-repulsion predicted with $U(2)$ could be absent due to potentially significant $U(2)$ symmetry breaking as is indicated in the phenomenology of EoS for dense baryonic matter. We will return to these matters in Secs.~\ref{repulsion} and \ref{rep core}.

\section{RG Analysis of the $\omega$-Nucleon Coupling}
\label{RG}

We have arrived at the DLFP (\ref{DLFP}) by linearizing dBHLS Lagrangian (\ref{dlfplag}) (which is gauge equivalent to nonlinear sigma model) in the mean-field approximation. Very near the fixed point, the resulting effective Lagrangian is a GML-type linear sigma model, to which the $\omega$ meson is minimally coupled. The isovector vector meson $\rho$ is decoupled in the limit.

In this section, we interpret the DLFP in terms of RG (renormalization-group) flow. To do this, we take the hidden local symmetry Lagrangian with baryons but without the dilaton $\chi$ (i.e., $\Slash{d}$BHLS).   As mentioned, in the chiral perturbation approach that we will be taking in this section, the role of scalar dilaton is problematic if introduced naively.
{We will first discuss the standard assignment for the nucleon, which gives a clearer picture of what's going on, and then consider the mirror assignment with parity-doubling.}
\subsection{Renormalization of the vector-nucleon coupling}
{In this subsection, we review shortly how to renormalize $g_{V\rho, \,\omega}$. The loop calculation is done in the background field gauge. When $\bar{\rho}^\mu$ and $\bar{\omega}^\mu$ are defined as the background fields of the $\rho$ and $\omega$ fields, the renormalization condition is given by
\begin{equation}
(g_{V\rho,\,\omega} -1)_{\rm bare} = Z_{3\rho,\,\omega}\left(Z_N Z_{\rho,\,\omega}^{1/2}\right)^{-1}(g_{V\rho,\,\omega} -1)\,, \label{ren}
\end{equation}where $Z_{3\rho,\,\omega}$ is given by the three point function of $\bar{\rho}(\bar{\omega})N_{+}N_{+}$ and
$Z_{\rho,\,\omega}$ presents the wavefuntion renormalization of $\bar{\rho}^\mu$($\bar{\omega}^\mu$) field.
Expanding $Z_{N,3\rho,3\omega} = 1 + Z_{N,3\rho,3\omega}^{(1)} \cdots$ and using $Z_{\rho,\,\omega} = 1$ for
the classical field $\bar{\rho}^\mu$($\bar{\omega}^\mu$), one obtains
the condition
\begin{equation}
(g_{V\rho,\,\omega} -1)_{\rm bare}
{}+ (g_{V\rho,\,\omega} -1)\left( - Z_{3\rho,\,\omega}^{(1)} + Z_N^{(1)} \right)
= \mbox{finite}\,,
\label{regular VNN}
\end{equation} where the superscript (1) represents one loop.
After taking the external momentum squared to be zero, the RGEs for $g_{V\rho,\,\omega}$ are obtained by taking the derivative of both side of Eq.~(\ref{regular VNN}) with respect to the (loop) momentum cutoff $\mu$. This is what is called in \cite{Shankar:1993pf} ``field theory approach'' to Wilsonian renormalization group . As we see from Eq.~(\ref{Nlagrangian2}), the $\rho NN$ coupling is distinguished from $\omega NN$ coupling in the Lagrangian when U(2) flavor symmetry is broken, and they are  renormalized differently from each other as in Eq.~(\ref{ren}) where $Z_{3\rho} \neq Z_{3\omega}$. But, when U(2) flavor symmetry is unbroken, {the $\rho NN$ coupling and the $\omega NN$ coupling carry one identical parameter} and hence there will be only one RGE for the coupling to nucleon.}

\subsection{RGEs in the Standard Assignment}\label{sectionrge}
{ In the standard assignment with $m_0=0$, we consider only} the positive-parity nucleon (set $N \equiv N_+$) in analyzing the RG properties of its coupling to mesons. The calculation is straightforward, the only difference from what was done in \cite{paeng} being that we have $SU(2)\times U(1)$ local symmetry instead of $U(2)$. Omitting the details that are given in \cite{paeng} -- apart from terms involving the $\omega$ meson, we simply write down the RGEs at one-loop order,
\begin{eqnarray}
\mu\frac{d m_N}{ d \mu} =
\frac{3 m_N}{16\pi^2} \tilde{\cal F}_m \,,\label{mn}
\end{eqnarray}
\begin{eqnarray}
\mu \frac{d}{d\mu} \left( 1- g_{V\rho} \right)
& = &
\frac{m_{N}^2}{8\pi^2F_\pi^2}\tilde{\mathcal F}_0
{}+ \left( 1-g_{V\rho} \right)\frac{1}{8\pi^2}\tilde{\mathcal F}_1
{}+ \left( g_{V\rho}-g_A^2 \right)\frac{1}{8\pi^2}\tilde{\mathcal F}_2\,,
\label{rgev:naive}
\\
\mu \frac{dg_A}{d\mu}
& = &
\frac{m_{N}^2g_A}{8\pi^2F_\pi^2}\tilde{\mathcal G}_0
{}+ \left( 1-g_{V\rho} \right)\frac{g_A}{8\pi^2}\tilde{\mathcal G}_1
{}+ \left( g_{V\rho}-g_A^2 \right)\frac{g_A}{8\pi^2}\tilde{\mathcal G}_2\,.
\label{rgea:naive}
\end{eqnarray}
\begin{eqnarray}
\mu \frac{d}{d\mu}  \left(g_{V\omega} -1\right)
& = &
{} 0\,.
\label{rgevNomega:naive}
\end{eqnarray}
The explicit expressions of $\tilde{\mathcal F}_i$ and $\tilde{\mathcal G}_i$
are given by
\begin{eqnarray}
\tilde{\cal F}_m
&=&
\frac{ g_{A}^2 }{ F_{\pi}^2} \left( \mu^{2} - m_N^2 \right)
{}- \frac{3}{2} \left(1 - g_{V\rho} \right)^{2} g_{\rho}^{2}
{}- \frac{1}{2} \left(g_{V\omega} -1\right)^{2} g_{\omega}^{2} \label{fm}
\\
\tilde{\mathcal F}_0
&=&
\frac{1}{4}\left( a_{\rho} g_A^2 + \frac{g_{V\rho}^2}{a_{\rho}} \right)\,,
\\
\tilde{\mathcal F}_1
&=&
\left[ \frac{g_A^2}{F_\pi^2} + \frac{g_{V\rho}\left(1+2g_{V\rho}\right)}{2F_{\sigma\rho}^2}
\right]\mu^2
{}- \frac{3}{2}\left( \frac{g_A^2}{F_\pi^2} + \frac{g_{V\rho}^2}{F_{\sigma\rho}^2}
\right)m_{N}^2
{}+ g_{\rho}^2\left( 2 - \frac{3}{2}g_{V\rho} \right)\,,
\\
\tilde{\mathcal F}_2
&=&
\frac{a_{\rho}}{2F_\pi^2}\mu^2\,,
\\
\tilde{\mathcal G}_0
&=&
\frac{1}{4}\left( g_A^2 + \frac{g_{V\rho}^2}{a_{\rho}} + 2g_{V\rho}\right)\,,
\\
\tilde{\mathcal G}_1
&=&
\left( \frac{2}{F_\pi^2} + \frac{1-g_{V\rho}}{F_{\sigma\rho}^2}\right)\mu^2
{}+ \frac{2g_{V\rho}}{F_{\sigma\rho}^2}m_{N}^2
{}- \frac{5}{2}a_{\rho}g_{\rho}^2\,,
\\
\tilde{\mathcal G}_2
&=&
-\frac{1}{F_\pi^2}\left( \mu^2 - 2m_{N}^2\right)\,.
\end{eqnarray}

To summarize the essential observations:
\begin{itemize}
\item While $(g_{V\rho}-1)=0$ coincides with the DLFP, $(g_{V\omega}-1)$ does not ``run" at one-loop order. There are two reasons for this. First, there are no contributions to RGE  from one-loop vertex corrections to the $\omega$-NN coupling because all the divergences in $\omega$NN 3-point functions are canceled by those in the nucleon self-energy diagrams figuring in wavefunction renormalization shown in Fig.~1\footnote{This result resembles the one-loop result in QED. In QED with $U(1)$ gauge symmetry, all divergent terms of the photon-electron-electron three point function are canceled by the divergent terms in the electron self energy diagrams. This cancelation is due to $U(1)$ gauge invariance. However this analogy does not extend to higher orders as mentioned in the text.}. Second,  there are no $\omega$ couplings to other mesons at tree order of the $\Slash{d}$BHLS Lagrangian, so there cannot be meson-loop contributions. However, as mentioned, at higher order this is no longer true. For instance there can be $\omega\rho\pi$ one-loop contribution to the $\omega$-nucleon coupling involving the homogeneous Wess-Zumino (hWZ) term in the anomalous part of the Lagrangian.  However the hWZ term goes as $\sim \left(\frac{p}{4 \pi F_\pi}\right)^4$, so the one-loop  contribution involving this vertex will correspond to normal two-loop order. Therefore  we expect $(g_{V\omega}-1)$ to run slowly,  if at all. It may be more appropriate to characterize it  as ``walking."
\item The RGE for nucleon mass, (\ref{mn}), has the fixed point $m_N=0$. But, if $\tilde{{\cal F}}_m <0$ at the fixed point, $m_N = 0$ cannot be an infrared fixed point.
Note that the $\omega$-nucleon coupling $\left(g_{V\omega} -1\right) g_\omega$  will affect the sign of $\tilde{{\cal F}}_m <0$. Since $\left(g_{V\omega} -1\right)$ does not run at one-loop level,  $\tilde{{\cal F}}_m$ could become negative at the fixed point unless $g_\omega = 0$ at that point. On the other hand, at two-loop or higher, $\left(g_{V\omega} -1\right)$ could be ``walking" to zero moving toward the DLFP, thereby making $\tilde{{\cal F}}_m >0$ near the fixed point, in which case $m_N=0$ will become an infrared fixed point. This indicates an intricate interplay between the nucleon mass and the $\omega$-nuclear dynamics. This is an important point to which we will return in Section \ref{rep core}.
\item {With the cutoff $\mu$ identified with a quantity related to density, as discussed in Appendix,  we see that $\tilde{\cal F}_m $ will tend to zero as density increases (or equivalently $\mu$ decreases). We consider the point where $\tilde{\cal F}_m \sim 0$  corresponds to the density $n \sim n_A$ because the nucleon mass given in RGE behaves similarly to the nucleon mass given in the mean field calculation in the vicinity of  $n \sim n_A$. Depending on the speed with which the $\omega$-NN coupling drops, the sign of $\tilde{\cal F}_m$, which controls the increase or decrease of the nucleon mass, will be determined. As a special case, if $\tilde{\cal F}_m $ stays near zero, the nucleon mass will then stay constant. This will correspond to the case of the dilaton condensate with B = 0.261. Thus, if the point where $\tilde{\cal F}_m \sim 0$ were the same as $n \sim n_A$, then DLFP would be reached at a density much higher than the saturation density since DLFP is characterized by $m_N$ going to zero.  It should however be stressed that even if the DLFP were far away from $n_A$, the physics above $n_A$ would be better controlled going toward the DLFP.}
\end{itemize}
%
\begin{figure}[h]
\begin{center}
\includegraphics[width=12.0cm]{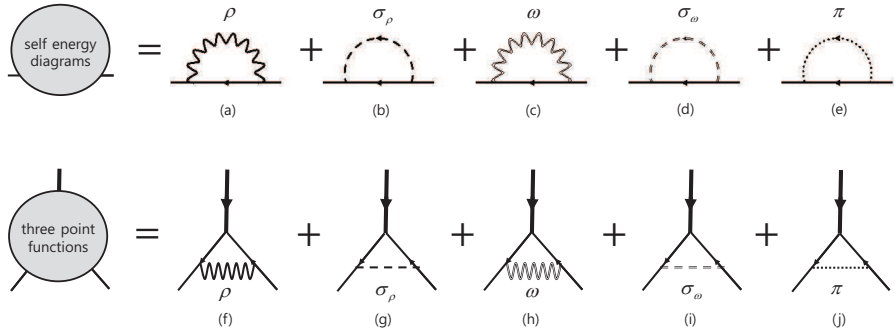}
\caption{The diagrams for RGE for $\omega$-NN coupling. The thin and thick solid line present the nucleon and the background field of the $\omega$ meson in the background field gauge respectively. The diagrams (a)-(e) contribute to the wave function renormalization of the nucleon field. In the calculation of $\omega$-NN coupling, the divergences given by the diagrams (f)-(j) are canceled by the divergences of the wave function renormalization of the nucleon field. }
\label{omega_nucleon diagram}
\end{center}
\end{figure}

\subsection{RGEs in the Mirror Assignment}
We now turn to nucleons in the mirror assignment, the parity-doublet model with $m_0\neq 0$. Here the treatment is a bit subtler because of non-zero $m_0$, which is chirally invariant and associated with the explicit breaking of scale invariance tied to the QCD trace anomaly. If $m_0\ll \Delta m_N$ where $\Delta m_N$ is the dynamically generated mass, then one could treat $m_0$ as a perturbation. In this case, we will have a qualitatively similar result to that of the baryon in the standard assignment. If on the other hand $m_0\gg \Delta m_N$, one could adopt heavy-baryon chiral perturbation theory. Both cases were discussed in \cite{paeng} for $U(2)$ symmetry. In between the two extreme regimes, we have no reliable tools to address the problem. In this paper, we are interested in the case where $m_0$ is ``big''\footnote{
In the vicinity of the chiral restoration point $n_c$, heavy-baryon ChPT could be applied even if $m_0$ were not so ``big." Since we are approaching the DLFP which is supposed to be close to $n_c$, our treatment should be reliable.}
 since that is what seems to be indicated in Nature.

As in \cite{paeng}, we will work to the leading order in $1/m_0$. With the heavy-baryon field defined as
\begin{equation}
\begin{pmatrix}
B_+
\\
B_-
\end{pmatrix}
=
\exp\left[ i m_0 v\cdot x \right]
\begin{pmatrix}
N_+
\\
N_-
\end{pmatrix}\,
\label{heavy}
\end{equation}
the Lagrangian (\ref{Nlagrangian2}) takes the form
\begin{eqnarray}
{\mathcal L}_N
&=&
i\bar{B}v^\mu D_\mu B - \Delta m_+ \bar{B}_+ B_+
{}- \Delta m_- \bar{B}_- B_-
\nonumber\\
&&
{}+ g_{V\rho}\bar{B}v^\mu\tilde{\alpha}_{\parallel\mu}B
{}+ g_{V\omega} \bar{B}v^\mu \left( \frac{ \partial_\mu \sigma_\omega}{2 F_{\sigma\omega}} - g_\omega \frac{\omega_\mu}{2}  \right)B
\nonumber\\
&&
{}+ g_A\bar{B}\left( 2S^\mu\rho_3 \tanh\delta
{}+ v^\mu\rho_1 \frac{1}{\cosh\delta} \right)
\tilde{\alpha}_{\perp\mu}B\,, \label{HBChPT}
\end{eqnarray}
where $S^\mu$ is the spin operator and
\begin{equation}
\Delta m_\pm = m_{N_\pm} - m_0\,.
\end{equation}
Making slight changes in \cite{paeng} for the local $SU(2)\times U(1)$ symmetry, we can easily write down the RGEs:
\begin{eqnarray}
\mu \frac{d M_{D} }{d\mu }
& = &
\frac{3g_{A}^2}{8 \pi^2 F_{\pi}^2 } M_{D} \left( \mu^{2} + 8 M_{D} \right) \left[ 1  + {\cal O} \left( \frac{1}{m_{0}^{2}}\right) \right] \,,
\label{rgemp0}\\
\mu \frac{d M_{S} }{d \mu } & = & 0\,,
\label{rgemm0} \\
\mu \frac{d}{d\mu} \left( 1-
g_{V\rho} \right)
& = &
\left[ \frac{M_{D}}{8\pi^{2} F_{\pi}^{2}} \bar{\mathcal F}_0  + \frac{\left( 1-g_{V\rho} \right)}{8\pi^{2}} \bar{\mathcal F}_1
{}+ \frac{\left( g_{V\rho}-g_A^2 \right)}{8\pi^{2}} \bar{\mathcal F}_2 \right] \left[ 1 + {\cal O} \left( \frac{1}{m_{0}}\right) \right]\,,
\label{rgev0}
\\
\mu \frac{dg_A}{d\mu}
 &= &
\left[ \frac{M_{D} g_{A} }{8\pi^{2} F_{\pi}^{2}} \bar{\mathcal G}_0  + \left( 1-g_{V\rho} \right) \frac{ g_{A}}{8\pi^{2}} \bar{\mathcal G}_1
{}+ \left( g_{V\rho}-g_A^2 \right) \frac{ g_{A} }{8\pi^{2}} \bar{\mathcal G}_2 \right] \left[ 1 + {\cal O} \left( \frac{1}{m_{0}}\right) \right]\,,
\label{rgea0}
\\
\mu \frac{d }{d\mu} \left(g_{V\omega} -1\right)
& = &
0\,,
\label{rgeomega0}
\end{eqnarray}
where $M_{S/D}=\left(m_{N_+}\pm m_{N_-}\right)^2/4$ and $\bar{\mathcal F}_i$ and $\bar{\mathcal G}_i$ are given by
\begin{eqnarray}
\bar{\mathcal F}_0
&=&
a_{\rho}g_A^2 + \frac{g_{V\rho}^2}{a_{\rho}}\,,
\\
\bar{\mathcal F}_1
&=&
\left[ \frac{g_A^2}{F_\pi^2} + \frac{g_{V\rho}\left(1+2g_{V\rho}\right)}{2F_{\sigma\rho}^2}
\right]\mu^2
{}+ 6 M_{D} \left( \frac{g_A^2}{F_\pi^2} + \frac{g_{V\rho}^2}{F_{\sigma\rho}^2} \right)
\nonumber\\
&&
{}- g_\rho^2\left( 4 - \frac{15}{2}g_{V\rho} + 3g_{V\rho}^2\right)\,,
\\
\bar{\mathcal F}_2
&=&
\frac{a_\rho}{2F_\pi^2}\mu^2\,,
\\
\bar{\mathcal G}_0
&=&
g_{A}^{2} + \frac{g_{{V\rho}}^{2}}{a_{\rho}}
{}+ 2 g_{{V\rho}}\,,
\\
\bar{\mathcal G}_1
&=&
\left( \frac{2}{F_\pi^2} + \frac{1-g_{V\rho}}{F_{\sigma\rho}^2}\right)\mu^2
{}- \frac{4g_{V\rho}}{F_{\sigma\rho}^2} M_{D}
\nonumber\\
&&
{}- g_\rho^2\left[ 3\left( 1 - g_{V\rho}\right) + \frac{5}{2} a_{\rho} \right]\,,
\\
\bar{\mathcal G}_2
&=&
-\frac{1}{F_\pi^2}\left(
\mu^2 + 4 M_{D}
\right)\,.
\end{eqnarray}

A quick glance at the RGEs in Eqs.~(\ref{rgemm0})-(\ref{rgeomega0}) gives us the infrared fixed point
\begin{eqnarray}
\left( M_S, M_D, 1-g_{V\rho}, g_{A} - g_{V\rho} \right) = \left( m_0^2, 0, 0, 0 \right)\,
\end{eqnarray}
which is the same as what was found in \cite{paeng}. This follows because of the absence of $\omega$-nucleon interaction terms for  RGEs at one-loop order.
%
 From the RGE of $\left(g_{V\omega} -1\right)$ in (\ref{rgeomega0}) follows that  the coupling $\left(g_{V\omega} -1\right)$ does not scale at one loop in this case also.  Similarly to the standard assignment, $\left(g_{V\omega} -1\right)$ will start scaling only at two-loop and higher-loop order.

\section{EoS in the Vicinity of the Dilaton-Limit Fixed Point}

Although, as stated in Introduction, the chiral restoration point $n_c$ may be inaccessible by local field theory, an interesting theoretic issue is what one can expect at very near the DLFP which is presumably close to the VM of hidden local symmetry. This could have phenomenological relevance because away from the DLFP, there could be precursor phenomena for which the treatment made in this paper is applicable.

In order to have the scaling properties deduced via RG flows given above make contact with nature, we need to translate the scale $\mu$ or cutoff $\Lambda/s$ (with $s$ the decimation going to $\infty$ in the Wilsonian sense) to density $n$.  In Appendix is given our reasoning. We cannot give a precise translation but we can say roughly that as density approaches the DLFP, $\mu$ should tend to zero. It is in this sense that the RG equations (\ref{mn}) - (\ref{rgevNomega:naive}) are to describe the flow to the infrared DLFP.

Two issues that can be addressed using the above translation are the nuclear symmetry energy of asymmetric nuclear matter and the repulsion induced by $\omega$ exchanges in dense matter. We briefly discuss these matters.
\subsection{Symmetry Energy}\label{SE}
The nuclear symmetry  $E_{sym}$ figures in the energy per baryon of asymmetric nuclear matter as
\be
E(n,\delta)=E_0(n)+E_{sym}\delta^2 + \cdots
\ee
where $\delta=(N-P)/(N+P)$ with $P(N)$ standing for the number of protons (neutrons) and  the ellipsis stands for terms ${\cal O}(\delta^4)$ and higher. We will treat the dBHLS Lagrangian in the mean-field (MF) approximation along the line developed in \cite{songetal,song}. Since the results are more or less the same for both assignments, we will treat only the case of standard assignment.

As mentioned before and explained in detail in \cite{songetal,song}, the MF approximation with chiral Lagrangians is justified up to -- and slightly above -- the nuclear saturation density which can be identified as the Landau Fermi-liquid fixed point. Given that we have no reliable guidance, we will simply extrapolate it to higher densities and see what can go wrong. From the mean field equations of motion for $\rho_3$, $\omega_0$ and $\chi$, one can immediately write down the potential energy contribution to the symmetry energy\footnote{It is generally agreed that the kinetic energy contribution is small compared with the potential energy term, particularly in the presence of strong tensor forces. We will not consider the kinetic energy term in this discussion.}
\be
E^{pot}_{sym}=\frac{{g^*_{\rho NN}}^2}{2{m^*_\rho}^2} n
=\frac{(g^*_{V\rho}-1)^2}{2(F^*_{\sigma\rho}\tilde{\chi}^*)^2} n\label{esym}
\ee
where we used
\be
g^*_{\rho NN} &=& (g^*_{V\rho} -1)g^*_\rho\,,\\
m^*_\rho &=& F^*_{\sigma\rho}\tilde{\chi}^* g_\rho^*\label{massp}
\ee
with { $\tilde{\chi}^*=\la\chi\ra^*/F_\chi$.}
Here and in what follows, the asterisk denotes density dependence.
Note that both the numerator and the denominator of (\ref{esym}) are sliding with density.

We should point out the difference between the mean field performed here (that we will call ``$\chi$MF") and the relativistic mean field found in the literature (that we will call ``standard MF"). In ``realistic" relativistic standard MF approaches, one adds higher dimension field operators based on ``naturalness conditions". With the parameters of the Lagrangian taken unscaling in density, without higher-dimension fields, the system at normal matter density is much too stiff, with a compression modulus more than 3 times what is indicated by nature. Here the mean field of  higher dimension field operators is to capture the physics of many-body forces and interactions and effectively introduces density dependence similar to what one has in $\chi$MF.  See for instance \cite{high-d-fields}. The higher-dimension field operators in the standard MF approaches modify the mass parameter (\ref{massp}) with additional density-dependent terms. But the shortcoming of these approaches is that there is no constraint that controls the EoS at higher densities.

In our $\chi$MF formulation with dBHLS Lagrangian also, we could in principle bring in higher-dimension field operators. However here, hidden local symmetry constrains the forms of the higher-dimension operators to include, so there is no arbitrariness as in standard MF. As recently stressed~\cite{dongetal}, the mean-field approach with hidden local symmetry  possesses correct thermodynamic constraints that allow a controlled extrapolation to higher density, such as, for example,  the rearrangement terms that reflect many-body terms, including certain many-body forces, that are captured in the high-dimension fields in the standard MF approach~\cite{song}.

The distinctive feature of the mean-field result with the dBHLS Lagrangian, Eq.~(\ref{esym}), if taken seriously at its face value, would imply that as density moves towards the DLFP, the symmetry energy should strongly decrease, going to zero at the fixed point. Such a softened symmetry energy, including the ``supersoft" one that drops to zero at $\sim 3n_0$, is one of the variety of scenarios found with suitable parameters picked in phenomenological models~\cite{supersoft}. We should mention that although somewhat extreme and perhaps exotic,  there is at present nothing that rules out such a soft symmetry energy.

An alternative scenario that is suggested by our results is that the expression (\ref{esym}) for the symmetry energy simply breaks down {\em before} reaching the DLFP.  When dense matter is described by having skyrmions embedded in a crystal lattice, the skyrmion matter representing normal baryonic matter transforms at a density $n_{1/2} \sim (1.3 - 2)n_0$ to a matter consisting of half-skyrmions. It involves the transition from a state $\la\bar{q}q\ra\neq 0$  to a state with $\la\bar{q}q\ra$ that goes to zero on the average in the unit cell. In the latter, the quark condensate is however non-zero locally, so there can be scalar density wave. Chiral symmetry is broken, however, with pions propagating, but with a higher-dimension order parameter.

This skyrmion-half-skyrmion  transition at $n_{1/2}$ turns out to have a drastic effect on the symmetry energy: The symmetry energy decreases in going toward $n_{1/2}$ as does the MF (\ref{esym}),  but then turns over at $n_{1/2}$ and increases strongly, making the $E_{sym}$ stiffer. This signals drastic departure from   (\ref{esym}).

It is easy to understand this ``cusp" structure in nuclear EFT with the dBHLS Lagrangian with the $(g_{V\rho}-1)$ coupling going toward the DLFP. In nuclear EFT approach,  the nuclear symmetry energy is dominated by the nuclear tensor forces that receive major contributions from one-pion exchange and one-$\rho$ exchange. It is roughly given by
\be
E^{pot}_{sym}\sim \frac{|V_T|^2}{\overline{E}}\label{efttensor}
\ee
where $V_T$ is the radial part of the tensor potential and $\overline{E}\approx 200$ MeV is the average excitation energy of the states dominantly excited by the tensor forces. The pion contribution and the $\rho$ contribution to the tensor forces, appearing with a different sign, tend to cancel as density increases. Thus we can see from (\ref{efttensor}) that the symmetry energy will decrease as the $\rho$ tensor gets enhanced by the dropping $\rho$  mass. However at $n_{1/2}$, the $\rho$ tensor coupling gets suppressed  by the decreas of the $(g_{V\rho}-1)$ coupling, thereby leaving the pion tensor fully active  above $n_{1/2}$. In \cite{dongetal}, this feature was exploited in describing the EoS of compact stars and succeeding to explain $\sim 2$ solar mass neutron stars. The significance of this result is that the mean-field result of Eq.~(\ref{esym}) cannot be extended to higher density when there is a topology change of the type we encounter, which is not encoded in the standard mean-field approaches.

{ As stated in the first section, the breakdown of mean-field theory at some density, revealed in our formulation, could imply the appearance of non-Fermi liquid structure. The equivalence of relativistic mean field theory to Landau Fermi-liquid fixed point theory which is believed to account for the success of RMF near nuclear matter density $n_0$ must therefore hold no longer in the phase approaching the DLFP. In terms of the topological structure exploited in \cite{dongetal} (see \cite{MRtalks} and references given therein and also \cite{maetalcrystal}), this implies that the half-skyrmion phase represents a non-Fermi liquid state. In condensed matter physics, such a topological phase change from Fermi liquid to non-Fermi liquid can be accessed by tuning physical parameters in a Hamiltonian (see, e.g., \cite{condensed}).  In our case we have no such luxury of being able to tune the parameters arbitrarily and measure the observables directly but it is plausible that density plays the crucial role.
}
\subsection{Nuclear Repulsion}\label{repulsion}
Contrary to the prediction made in \cite{sasaki} with $U(2)$ symmetry for the $\rho$ and $\omega$,  the nuclear phenomenology of \cite{dongetal} shows that the strong suppression of the $\omega$-induced repulsion cannot take place as one approaches the DLFP. One can see this also in the $\chi$MF with the dBHLS Lagrangian. The potential energy per baryon given by an $\omega$ exchange between two nucleons is of the form
\be
E^{pot}_\omega= \frac{{g^*_{\omega NN}}^2}{2{m^*_\omega}^2}n +\cdots
\ee
where the ellipsis stands for three-body and higher-body contributions which will be  of $(\frac{{g^*_{\omega NN}}^2}{2{m^*_\omega}^2})$ or higher order. Using the tree-order mass formula for the in-medium $\omega$, ${m^*_\omega}^2=(g^*_\omega F^*_{\sigma\omega} \tilde{\chi}^*)^2$
and the coupling $g^*_{\omega NN}=g^*_\omega (g^*_{V\omega}-1)$,
we have that
\be
E^{pot}_\omega= \frac{(g^*_{V\omega}-1)^2}{2 (F^*_{\sigma\omega} \tilde{\chi}^*)^2}n +\cdots
\ee
We expect that while the numerator will drop only slowly -- if at all,  the denominator will drop somewhat less slowly.
 Thus one sees that the $\omega$ repulsion will not undergo an appreciable change as the DLFP is approached.

\subsection{A Mean-Field Model}
\label{mf}

The subtle interplay between the nucleon mass and the $\omega$-nucleon
coupling can be seen in the mean-field calculation with dBHLS Lagrangian
of Section \ref{dBHLS}. Below we examine the standard scenario with $m_0=0$.
The dilaton potential is taken in the standard form~\cite{schechter},
\begin{equation}
V(\chi)
= - \frac{m_\chi^2}{8 F_\chi^2}
\left[ \frac{1}{2}\chi^4
{}- \chi^4\ln\left( \frac{\chi^2}{F_\chi^2}\right)\right]\,.
\end{equation}
Following \cite{Song:1997kn} the thermodynamic potential is constructed
with a density-dependent $\omega$-nucleon coupling,
\begin{eqnarray}
\Omega(\chi,\, n)
&=&
\frac{1}{4\pi^{2}} \left[ 2 E_{F}^{3} p_{F} - m_{N}^{\ast 2} E_{F} p_{F}
{}- m_{N}^{\ast 4} \ln \left( \frac{E_{F} + p_{F} }
{m_{N}^{\ast}} \right) \right]
{} + \frac{\left(g_{V\omega}^\ast -1 \right)^2}
{2F_{\sigma\omega}^2 \chi^2/F_{\chi}^2} n^2
\nonumber\\
&&
{} -  \frac{ m_\chi^2}{8} F_{\chi}^{2}
\left\{\left( \frac{\chi^2}{F_\chi^2}\right)^{2}
\left[ \frac{1}{2} - \ln\left(\frac{\chi^2}{F_\chi^2}\right)
\right] - \frac{1}{2} \right\} -\mu(n) n\,,
\label{omega}
\end{eqnarray}
where $E_F = \sqrt{p_F^2 + m_N^{\ast\, 2}}$
and the chemical potential is given as a function of density $n$ by
\begin{equation}
\mu(n)
= E_F(n) + \frac{\left( g_{V\omega}^\ast -1\right)^2}
{F_{\sigma\omega}^2 \chi^2/ F_\chi^2} n
{}+ \frac{\left( g_{V\omega}^\ast -1\right)}
{F_{\sigma\omega}^2 \chi^2/ F_\chi^2}n^2
\frac{\partial \left( g_{V\omega}^\ast -1\right) }{\partial n}\,.
\end{equation}
The nucleon mass is connected to the $\omega$-nucleon coupling by the
equation of the motion for $\chi$ and $\omega$, and the in-medium property
of the $\chi$ condensate -- equivalently the in-medium mass of the dilaton
-- controls
the behavior of the nucleon mass at high density. The nucleon mass depends
on $\bchi=\la \chi\ra$ via
\begin{equation}
m_N^\ast = g\bchi\,.
\end{equation}
The gap equation for $\chi$ is found as
\begin{equation}
\chi\left[
\frac{ m_N^2}{\pi^2 F_\chi^2}\left(
p_F E_F - m_N^{\ast\, 2}\ln\left(\frac{p_F + E_F}{m_N^\ast}\right)
\right)
{}- \frac{\left(g_{V\omega}^\ast -1 \right)^2}
{F_{\sigma\omega}^2 \chi^4/F_{\chi}^2} n^2
{}+ \frac{ m_\chi^2}{2} \left( \frac{\chi^2}{F_\chi^2} \right)
\ln\left(\frac{\chi^2}{F_\chi^2}\right)
\right]=0\,.\label{gapchi}
\end{equation}
In the mean field approach, taking the limit $\mbox{tr}\left[\Sigma
\Sigma^\dagger \right] \rightarrow 0$ is replaced by $\bchi \rightarrow 0$.
The slowly decreasing $\omega$-nucleon coupling not only causes the nucleon
mass to drop slowly,
but also delays the dilaton limit, $g_{A} = g_{V\rho}$, to higher density.

\begin{figure}[h]
\begin{center}
\includegraphics[width=10cm]{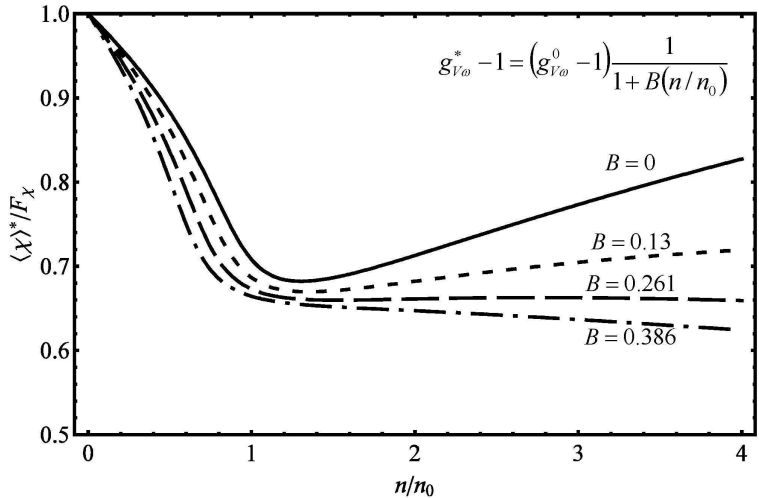}
\caption{The ratio $m_N^*/m_N\approx \la\chi\ra^*/\la\chi\ra_0$  as a function of
density for varying density dependence of $g_{V\omega}^\ast$.
What is notable is that the nucleon mass stops dropping at a density slightly
above nuclear matter density $n_0$ and stays more or less constant above that
density.
}
\label{mass_coupling}
\end{center}
\end{figure}

 In Fig.~\ref{mass_coupling} is shown an illustration of the above
observation introducing a simple parametrization\footnote{We find that the coupling
$g_{V\omega}-1$ does not scale at one loop. Scale dependence  starts to
show up at two-loop level and it is expected to decrease the coupling as the dilaton limit fixed point is approached. Thus, at any finite-loop order, the coupling could
have a rather weak scale-dependence. As mentioned in Appendix, we have decoded the scale dependence into a density dependence with the in-medium coupling represented in a weakly decreasing function of the form Eq.~(\ref{omega_coupling}).},
\begin{equation}
\frac{g_{V\omega}^\ast - 1}{g_{V\omega}-1}
= \frac{1}{1 + B n/n_0}\,. \label{omega_coupling}
\end{equation}
For a given constant $B$, the nucleon mass is calculated by fitting the
binding energy and the pressure of nuclear matter at $n_0$.\footnote{
 The two density-dependent quantities involved are $m_\chi^*$ and
$g_{V\omega}^*$ that are determined by the binding energy and the pressure
at $n=n_0$ for given $B$. The behavior of the nucleon mass of
Fig.~\ref{mass_coupling} then follows.} It is remarkable that the nucleon
mass drops\footnote{Up to the normal nuclear matter density, the dilaton condensate is related to the quark condensate. The $\sim 30\%$ drop of the dilaton condensate at $n_0$ from the vacuum value predicted here  is consistent with the empirical value of the quark condensate estimated from the in-medium pion decay constant measured in deeply bound pionic states.} almost linearly in density up to slightly above $n\sim n_0$, say,
$n_{A}$\footnote{In the text, this was referred to as $\tilde{n}$.
We conjectured there that it corresponds to the skyrmion-half-skyrmion
transition density $n_{1/2}$},  and then stays more or less constant.

{ One finds that the behavior of the dilaton condensate drastically changes at $n_A$. This  behavior is caused by the change of the solution for Eq.~(\ref{gapchi}) as
 \begin{equation}
 \bar{\chi}:\, \bar{\chi}_+ \rightarrow \bar{\chi}_-
 \end{equation}
 at $n \sim n_A$, where $\bar{\chi}_{\pm}$ are given by
\begin{equation}
\frac{m_\chi^2}{F_{\chi}^2}\left|\ln\left(\frac{\bchi_{\pm}^2}{F_\chi^2}\right)\right|\bchi^3_{\pm} = \frac{3 m_N}{4 F_\chi} n \pm n\sqrt{\left(\frac{3 m_N}{4 F_\chi}\right)^2 -2\frac{m_\chi^2}{F_{\sigma\omega}^2}
\left|\ln\left(\frac{\bchi_{\pm}^2}{F_\chi^2}\right)\right|\left( g_{V\omega}^\ast - 1\right)^2}\,\label{EoM3}
\end{equation}
that are solutions to Eq.~(\ref{gapchi}) in the approximation that $p_F / m_N$ is small. The drastic change occurs at $n \sim n_A$, where $\bar{\chi}_+ = \bar{\chi}_-$ and the quantity $\bar{\chi}$ follows the behavior of $\bar{\chi}_-$ after $n \sim n_A$. }

One can readily understand the above interplay between the nucleon mass and the $\omega$-NN coupling after $n \sim n_A$. The behavior of $\bchi$ depends on how the product $(g_{V\omega}^\ast-1)^2 n^2$ goes with density. { If we expand the solution $\bar{\chi}_-$ in terms of $R(n)$ defined as
\begin{equation}
R(n) \equiv 2 \frac{m_\chi^2}{F_{\sigma\omega}^2} \ln \left( \frac{F_\chi}{\bar{\chi}_-} \right)^2 \left( g_{V\omega}^\ast - 1 \right)^2/\left( \frac{3m_N}{4F_\chi}\right)^2\,,
\end{equation}
$\bar{\chi}_-$ is simplified to
\begin{equation}
\bchi_-^3 \sim \frac{4}{3}\frac{F_\chi^3}{F_{\sigma\omega}^2 m_N}
\left(g_{V\omega}^\ast -1\right)^2 n + {\cal O}\left( R(n) \right)\,
\end{equation} at intermediate density, $n > n_A$, where $R(n) < 1$.}
Consequently, if $g_{V\omega}^\ast$ is constant, i.e. $B=0$, the VEV goes like $\bchi \sim n^{1/3}$. Whereas when the effective coupling varies with density as
$(g_{V\omega}^\ast-1)^2 \sim 1/n$, one finds $\bchi \sim$ const. as
well captured in Fig.~\ref{mass_coupling}.

As stressed we do not expect the DLFP to be on top of chiral restoration (in the chiral limit), but it may be close to it. So an interesting question is whether our mean field model can say something about the chiral restoration transition.

With the conformal compensator prescription, the $\omega$-meson mass term in the present context carries $\chi^2$. This appears in the mean-field thermodynamic
potential~(\ref{omega}) in the form of $(g_{V\omega}^\ast -1)^2 n^2/\chi^2$
by use of the equation of motion for $\omega_0$. Once the density is turned
on, the inverse power of $\chi$ generates a divergent contribution to the
entire $\Omega$ peaked at $\chi=0$. This huge barrier prevents the VEV $\bchi$
from approaching the scale-symmetry (or equivalently chiral-symmetry) restored state, $\bchi= 0$. This suggests that the $\omega$-meson mass may not be associated entirely with the spontaneously broken scale symmetry. As discussed in \cite{LR}, an explicit scale symmetry breaking is needed to trigger spontaneous symmetry breaking associated with the dilaton $\chi$.  For this, an additional scalar field, denoted $\chi_h$ in \cite{LR}, should figure. How it should be introduced in the formulation of this paper is not clear. Just to have an idea of its potential role, let us consider that the scalar has two components $(\chi_s, \chi_h)$ and identify $\chi_s$ with the dilaton we have been using. Let us assume that we have a linear combination of the two scalars. We will then have
$(g_{V\omega}^\ast -1)^2n^2/(a\chi_s + b\chi_h)^2$ in $\Omega$ which now stays
finite at $\chi_s=0$ since the VEV of $\chi_h$ is non-vanishing. The height of
the potential barrier is governed by the composition of $\chi_s$ and $\chi_h$ in the $\omega$ mass. When the soft dilaton plays a minor role there,
$\bchi$ turns out not to stay constant any more but decreases monotonically. In order to have the flat $\chi$ after the onset density $n_A$ up to some density $n_B$ and then have it drop to zero for chiral restoration, some sort of level crossing must take place between the $\chi$'s as density is increased. This may be related to the still-open problem of low-mass scalars in nuclear and hadron physics vis-\`a-vis with $f_0(500)$.

Here we should stress that this mean-field calculation was made in the standard
assignment, $m_0=0$ in the beginning. Nevertheless, we have found
$m_N^\ast \sim 0.7$ in high density, indicating that a non-vanishing $m_0$
emerges dynamically.
The interplay between the nucleon mass and the $\omega$-nucleon coupling
as revealed in this way is uncannily similar to what was found by the RGE
analysis given in this paper and consistent with what was phenomenologically
observed in nuclear EFT description with a BR scaling modified by topology
change~\cite{dongetal}. (See also \cite{maetalcrystal}.)

\section{Conclusions}
\label{rep core}
Although we have neither mathematically rigorous arguments nor any results of phenomenological studies using the formulation of \cite{dongetal}, we can still make a qualitative connection between the main source (or origin of ) of nucleon mass and the property of the $\omega$-nucleon coupling in medium. As discussed recently~\cite{MRtalks},  there are at least two reasons to think that the nucleon mass has a large component that is chiral-scalar and hence does not disappear at the chiral transition. One is a lattice result: By artificially unbreaking chiral symmetry in a dynamical lattice simulation~\cite{Lang-unbreaking} (see also \cite{suganuma}), Glozman et al. find that baryons -- and also mesons -- remain massive after chiral symmetry is supposedly restored~\cite{glozman}. In fact, phrased in the form of Eq.~(\ref{mass}), $m_0$ is found to be surprisingly large. The other is evidence from dense skyrmion matter where one notes that as density increases, the energy per baryon of the system scales with the effective pion decay constant that falls quite slowly once the system enters into the half-skrymion phase~\cite{MRtalks}.

As proposed in \cite{dongetal,MRtalks} and stated above, the presence of a large chirally invariant term in the nucleon mass can be effectively captured in the parity-doublet baryon structure with the nucleon mass given by (\ref{mass}), $m_N=m_0 +\Delta m (\kappa)$.
In principle, one could determine $m_0$ from low-energy hadronic processes involving parity-doublet baryons with  $\Slash{d}$BHLS. For this, high-order chiral perturbation theory with the vector mesons suitably incorporated would be needed. Such a formulation is not yet available.

A rough range available up to date is that $m_0$ could range from $\sim 200$ MeV to $\sim 800$ MeV~\cite{sasaki,paeng,rangem0}.

Let us briefly discuss what we can say from nuclear phenomenology with, say, $m_0\approx 800$ MeV.  In \cite{dongetal}, the structure of dense skyrmion matter simulated on crystal lattice treated with a truncated form of dBHLS~\cite{LPR} was translated into the scaling behavior of the parameters figuring in the dBHLS Lagranigian. The rationale for such a procedure was that the topology change from skyrmions to half-skyrmions present in the crystal matter could be interpreted as a change in the parameters as density increases across $n_{1/2}$. For the nucleon mass, the scaling was taken to follow the empirical trend up to the normal matter density $n_0$, giving $m^*_N/m_N\sim 0.8$ at $n=n_0$, and then assumed to remain unchanged after $n_{1/2}\sim 1.2n_0$. We can consider this choice as corresponding to $m_0/m_N\sim 0.8$ in the parity-doublet dBHLS model with $\Delta m\rightarrow 0$ as $n\rightarrow n_{1/2}$ at which $\la\bar{q}q\ra^*$ vanishes\footnote{Although the condensate $\la\bar{q}q\ra$ is not a proper order parameter for chiral symmetry for the crystal matter, we will identify it to be associated with the dynamically generated mass term $\Delta m$}. For this value of $m_N^*$ -- and with other parameters fixed, it was found that even a slight decrease in the $\omega$-nucleon coupling would generate instability of nuclear matter for $n> n_{1/2}$. This is an indication that for the given value of $m_N^*$, the $g_{\omega NN}$ cannot run much in density after $n_{1/2}$.

The reason for this phenomenon is easy to understand. In nonrelativistic description in nuclear EFT, lowering $m_N^*$ engenders strong suppression in attraction, the suppression factor going like $\sim (m^*_N/m_N)^2$, basically a relativistic effect on the scalar density at higher baryon density. On the other hand, lowering $g^*_{\omega NN}$ decreases, roughly quadratically,  the $\omega$-exchange repulsion, causing instability of the matter at high density. Therefore with other parameters fixed, the two are tightly correlated in the EoS. Given that $m_0$ is not known at present, one may vary both $m_0$ and $g^*_{\omega NN}$ in such a way that the EoS remains within the ranges provided by heavy-ion experimental data available up to $\sim 4 n_0$ and consistent with neutron-star observables. This could offer some indication as to whether $(g_{V\omega}-1)$ runs or rather walks as we found in this paper. This feature is illustrated by a simple mean-field treatment with dBHLS as given in Sec.~\ref{mf}. It should be emphasized that the $m_0$ is the part of mass that has nothing direct to do with the spontaneous breaking of chiral symmetry, so if it is as big as 80\% of the nucleon mass, the origin of the bulk of proton mass is yet to be explained~\cite{MRtalks}.

\section*{Addendum}
{ After this work was posted on arXiv,  a calculation of dense matter by putting skyrmions on crystal with HLS Lagrangian with all the coefficients of the $O(p^4)$ terms of the Lagranigian (including the homogeneous Wess-Zumino terms) fixed from holographic QCD~\cite{maetal} was completed~\cite{maetalcrystal}. The results agree with what's obtained in this paper.}

\subsection*{Acknowledgments}
We are grateful for correspondence with Tom Kuo on the role of new BR scaling in the EoS of dense matter and for discussions with Masayasu Harada and Yong-Liang Ma on dilatons. We acknowledge the hospitality at APCTP where part of this work was
done. This work was partially supported by the WCU project of the Korean Ministry of Educational Science and Technology (R33-2008-000-10087-0). The work of C.S. has been partly supported by the Hessian LOEWE initiative through the Helmholtz International Center for FAIR (HIC for FAIR).

\newpage

\appendix

\centerline{\large\bf  APPENDIX}

\setcounter{section}{0}
\renewcommand{\thesection}{\Alph{section}}
\setcounter{equation}{0}
\renewcommand{\theequation}{A.\arabic{equation}}
\section*{The interpretation of the cutoff relating to the density}

{ In this appendix, we present a far-from-rigorous argument to relate the scale $\mu$ in RGEs to  density. In order to make a consistent and realistic treatment of dense matter with an effective field theory Lagrangian of the type we are considering, one would first have to arrive at a Fermi sea and then obtain a Fermi-liquid fixed point theory to correctly describe physics of normal nuclear matter. One may then extrapolate the theory to higher density assuming that the Fermi-liquid structure continues to be valid. Here we are not in a position to do so.  What we will do here is simply to assume that the Fermi-sea structure is arrived at with our effective field theory and interpret the RG analysis we have made in the paper in terms of the kinematics appropriate to that of the Fermi sphere.

We adopt the basic premise of Wilsonian  RG to integrate out higher modes and arrive at the Fermi liquid fixed point structure as discussed in  \cite{Shankar:1993pf}.   We will assume as in Walecka mean field theory~\cite{matsui} that relativistic mean field theory with our dBHLS Lagrangian gives a Fermi-liquid fixed point theory at the saturation density $n_0$. Suppose the  effective (dBHLS) Lagrangian that we start with has a set of parameters, $a_i$, and fields, $\phi_i$
\be
{\cal L} = {\cal L}(\phi_i, a_j).
\ee
Using the kinematics defined in the Fermi sphere as given in Fig.~\ref{interpret},  we define the integration momentum $k$
\begin{equation}
k = |K| - K_F\,,
\end{equation}
and write the action as
\begin{equation}
S = \int^{\Lambda}_{-\Lambda} {\cal L}(\phi_i, a_j) \frac{d^4 k}{(2\pi)^4}\,,
\end{equation}
where $K_F$ is the Fermi momentum and $K$ is the momentum of the particle in medium satisfying the dispersion relation $E = \frac{K^2}{2m}$. Then $k$ is the momentum component normal to the Fermi surface with the cutoff, $\Lambda$, that defines the scale up to which the effective theory is valid.
\begin{figure}[h]
\begin{center}
\includegraphics[width=10cm]{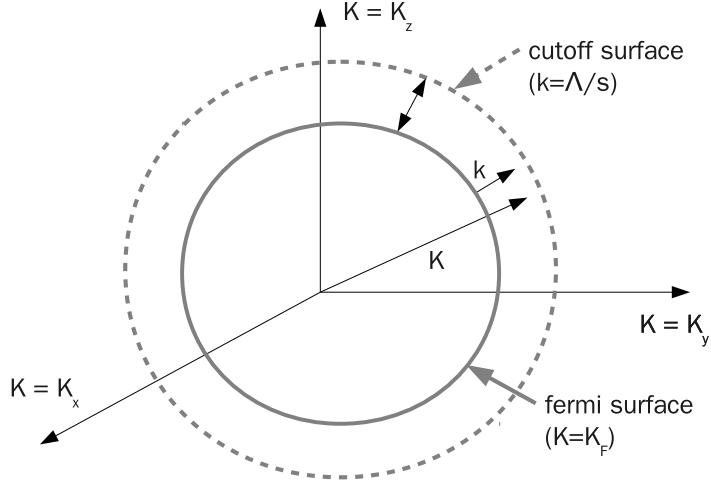}
\caption{The interpretation of k, K and the cutoff in the momentum space.
}
\label{interpret}
\end{center}
\end{figure}
When the fields are separated into high and low modes as
\begin{eqnarray}
\phi_{iL} &=& \phi_i \quad \textnormal{for } 0 < k < \Lambda/s\,,\\
\phi_{iH} &=& \phi_i \quad \textnormal{for } \Lambda/s < k < \Lambda\,,
\end{eqnarray}
the action is given by
\begin{equation}
S(\phi_{iL}, \, \phi_{iH}) = S_0( \phi_{iL} ) + S_0( \phi_{iH} ) + S_I( \phi_{iL} , \,\phi_{iH}).
\end{equation}
After integrating out the high modes -- $\phi_{iH}$ -- we get
\begin{eqnarray}
Z
&=&
{} \int \prod_j \left[ d \phi_{jH}\right] \left[ d \phi_{jL}\right] {\rm e}^{-\left\{ S_0 [\phi_{iL} ] + S_0 [\phi_{iH}] + S_I [\phi_{iL},\, \phi_{iH} ] \right\}}\,,\\
&=&
{} \int \prod_j \left[ d \phi_{jL}\right] {\rm e}^{ S_{eff} [\phi_{iL} ] }\,,
\end{eqnarray} where
\begin{equation}
S_{eff} [\phi_{iL}] = \ln \left\{ {\rm e}^{- S_0 [\phi_{iL} ]} \int \prod_j \left[ d \phi_{jH}\right]  {\rm e}^{-\left\{ S_0 [\phi_{iH}] + S_I [\phi_{iL},\, \phi_{iH} ] \right\}} \right\}\,.
\end{equation}
The resulting effective action is dependent on the cutoff $\Lambda/s$ with $s$ running from $1$ to $\infty$.


 We now wish to relate the dropping cutoff, $\Lambda/s$, to the increasing density in Figure~\ref{interpret}. For this, let us suppose that we are going to the region where $\Lambda/s$ goes to zero with $K_F$ fixed. This is described in Figure~\ref{interpret} as the ``cutoff surface," $k = \Lambda/s$, approaching the Fermi surface, $K = K_F$. The parameters will change as the cutoff, $\Lambda/s$, is changed. The interval between the Fermi surface and the cutoff surface will then get reduced.

 It is more convenient for our purpose to interpret the above process in a different way.
Let $K_F$ increase with $\Lambda/s$ fixed, that is,  let the Fermi surface approach the cutoff surface, so that the interval between the Fermi surface and the cutoff surface goes to zero. Then, the magnitude of the possible maximum momentum of $k$ tends to zero which  is effectively the same as the cutoff, $\Lambda/s$, going to zero.  We approach the infrared region, $\Lambda/s \rightarrow 0$, by increasing $K_F$.

In our RGEs given in the text, the $\mu$ is interpreted as the cutoff
in Wilsonian RG approach,
\begin{equation}
\mu = \Lambda/s\,.
\end{equation}
If our argument relating the cutoff to density can be applied to our system, we can then infer a rough relation between $\mu$ and $K_F$ in the form
\begin{equation}
\mu_0 - \mu = K_F\,, \label{mutodensity}
\end{equation} where $\mu_0 \equiv \mu(n=0) = \Lambda$ for $s=1$.
This shows that as $\mu$ decreases $K_F$ increases. If our interpretation of  Eq.~(\ref{mutodensity}) is correct, then we can say that the physical quantities flowing with RGEs in our paper text capture the medium effect, and flow to DLFP as density increases because DLFP is evaluated as an IR fixed point. The slowly decreasing behavior of $g_{V\omega} -1$ given in RGE is applied in the mean field calculation to the dilaton condensate, and gives the interplay between the nucleon mass and the $\omega$-nucleon coupling in $n > n_A$.}

\end{document}